\documentstyle[aps,pre,preprint]{revtex}

\begin{document}
\tighten
\draft

\title{Topological aspects of chaotic scattering in higher dimensions}

\author{Zolt\'an Kov\'acs}
\address{Institute for Theoretical Physics, E\"otv\"os University \\ 
         Pf.\ 32, H--1518 Budapest, Hungary}
\author{Laurent Wiesenfeld}
\address{Laboratoire de Spectrom\'etrie Physique, 
                        Universit\'e Joseph-Fourier-Grenoble \\ 
         BP 87, F--38402 Saint-Martin-d'H\`eres Cedex, France}

\date{30 September 1998}

\maketitle

\begin{abstract}
We investigate the topological properties of invariant sets 
associated with the dynamics of scattering systems with three 
or more degrees of freedom. 
We show that the asymptotic 
separation of one degree of freedom from the rest 
in the asymptotic regime, a common property in a large class
of scattering models, 
defines a dynamical object with phase space separating invariant 
manifolds and an invariant set with larger dimension than 
that of the set defined by bounded orbits.
In particular, the set of typical periodic orbits involving all the
degrees of freedom of the system form 
a nowhere dense subset of the large invariant set.
\end{abstract}

\pacs{05.45.+b, 34.10.+x}

\section{Introduction}

Chaotic scattering in open Hamiltonian systems with two 
degrees of freedom has become a well understood phenomenon \cite{r1}.
A central role is played in these processes by the 
{\em chaotic invariant set} which consists of all the bounded
orbits, i.e.\ those without the simple incoming and
outgoing asymptotic motions characteristic of the system.
The invariant set is globally hyperbolic which means that 
its orbits can attract incoming trajectories along their
stable manifolds extending smoothly towards the asymptotic
region. 
The trajectories are then forced into a temporarily chaotic behaviour
in the vicinity of the chaotic invariant set, 
before eventually escaping along the
unstable manifold of the set.
For initial conditions exactly on a stable manifold branch, 
the trajectories may even become asymptotically trapped
by the corresponding bounded orbit; this leads to 
singularities on a fractal set in the scattering functions.

In the two-dimensional Poincar\'e sections generally used for the
representation of the dynamics, bounded orbits appear as
points forming a double Cantor set structure. 
Stable and unstable manifolds of the bounded orbits 
show up as continuous lines.
The stable manifold lines are of codimension one in these cases,
so they can separate the phase space into distinct regions
corresponding to qualitatively different scattering
trajectories. 
The same property makes it also possible to capture the
singularities associated with the irregular nature of 
scattering in any generic one-dimensional family of initial
conditions. 

There are two traditional ways to look at the chaotic
invariant set \cite{r1}. 
(i) One is to consider a simple periodic orbit that plays a
basic role in organizing the scattering process: 
usually this periodic orbit lies at the edge of the scattering region, 
and trajectories approaching it escape or do not escape 
the interaction region
depending on which side of its 
stable manifold they are while approaching. 
We will call such objects {\em gates} 
in the following \cite{r2,comment}; as we will show, they can be 
more complicated than periodic orbits.
The set of all the homoclinic points
(intersections of the stable and unstable manifolds) 
consists of trajectories doubly asymptotic to the gate.
The chaotic invariant set is then obtained as the closure
(in a mathematical sense) of the set of the intersection
points and contains also {\em all} the possible periodic
orbits of the system. 
(ii) The alternative approach considers the chaotic invariant
set as the closure of the set of all the periodic orbits;
the closure in this case generates trajectories doubly
asymptotic to periodic ones.
These two ways give equivalent results in systems with 
two degrees of freedom (apart from pathological
exceptions), and many studies have made use of this duality
in the analysis of such chaotic scattering processes.

In fact, it has become an article of faith that all chaotic
invariant sets, whether related to transient or permanent chaos,
can be considered as the collection of all the periodic orbits
plus the limiting orbits provided by the closure process.
In this paper, we show that this assumption is to be taken 
with caution
in a large class of scattering systems with at least 
three degrees of freedom.
There exist more than one
invariant sets, with different fractal dimensions, 
obtained in different constructions. 
Moreover, as far as the escape process is
concerned, typical periodic orbits involving 
all the degrees of freedom of the system 
play a marginal role compared to the 
set generated by the gate. 

This difference appears whenever the gate is a
codimension one invariant subset in configuration space.
A typical situation resulting in such subsets in scattering systems 
is the separation of one translational degree of freedom 
from internal degrees of freedom in the asymptotic region.
We will demonstrate this point in a simple planar 
three-body scattering
problem describing the reactive collision of a single atom with
a diatomic molecule.
We explain our findings using this model as an illustrative example
that we analyze from the point of view of
treating general scattering processes.

\section{A simple model}

A typical three-body scattering problem appearing in many contexts 
is a single particle interacting with a two-body system 
(usually integrable in itself); 
the most common examples are from chemical physics (reactive
collisions) and celestial mechanics (planetary motion). 
Model problems of this type have been studied in several varieties,
including ones restricted to two degrees of freedom.
In fact,  problems like reactive collisions in 
collinear or T-shape configurations \cite{react,r2} 
or the interaction of satellites in 
coplanar circular orbits (Hill's problem) \cite{hill} 
have become well-known representatives of chaotic scattering.

However, a more general treatment involving more than two degrees of
freedom in these systems (and similar ones) is clearly necessary.
We have chosen a planar atom-diatom collision with pairwise 
Morse potentials between the atoms at a total energy below the
dissociation threshold 
so scattering consists of an exchange reaction.
There are three qualitatively different channels corresponding
to which of the three atoms escapes 
after the decay of the transient complex.
Since we are more interested in the general topological properties of
chaotic scattering rather than in accurate modelling of particular
chemical reactions, 
we have chosen identical atomic masses and Morse potential parameters
for simplicity.

In a center-of-mass coordinate system, the total number of degrees of
freedom of this planar system is four, 
with total energy $E$ and angular momentum $J$ conserved. 
By choosing suitable  new coordinates,  
the angle variable conjugate to $J$ can be
separated from the other three that still contain all the information
on the relative distances of the atoms.
For our purposes,
the most appropriate choice is an abstract representation 
of the configuration by Cartesian coordinates $(x,y,z)$ based on 
hyperspherical coordinates widely used for three-body problems 
\cite{hyper1,hyper2,hyper3}. 

The new variables and their time derivatives can then be used to express the
kinetic energy $K$ \cite{hyper2} and the interatomic distances 
$r_{12}$, $r_{13}$ and $r_{23}$ \cite{hyper3}.
The actual expressions are rather complicated; we refer the interested
reader to the literature on these coordinate systems, especially the 
references given above.
The Hamiltonian takes the (dimensionless) form
\begin{equation}
     H(x,y,z,p_x,p_y,p_z) = K + V_M (r_{12}) + V_M (r_{13}) + V_M (r_{23})
\end{equation}
where $V_M (r) = (1- e^{-r})^2$ is the Morse potential.
For simplicity, we confine ourselves to cases with $J=0$.
The explicit Hamiltonian and the equations of motion are given in 
Eqs.~(60--66) of Ref.~\cite{hyper2}.

Important dynamical symmetries and special cases are reflected
in these coordinates.
The potential is mirror symmetric with respect to the plane $z=0$ which 
corresponds to collinear configurations, and 
the plane $y=0$ together with its two images obtained by rotation around
the $z$ axis by $\pm 2\pi /3$ contain symmetric T-shape
configurations.
For a given energy $E$ below the total dissociation threshold 
$E_{\rm tot}=3$,
the $z$ variable is confined to a range 
$[-z_{\rm max}, z_{\rm max}]$, while $x$ and $y$ can be arbitrarily large.
The three channels corresponding to the three different outcomes 
of the reaction extend to infinity along the intersection lines of
the T-shape planes with the collinear one.
The boundaries of the three-dimensional energy surface,
defined by $K=0$,
are plotted as usual surfaces in the abstract $(x,y,z)$ space 
for a typical scattering energy in Fig.~\ref{fig-cont}.

The escape channels of the scattering have asymptotically an
axial symmetry as the distance from the origin tends to
infinity. 
This is due to the separation of the translational motion of the 
outgoing single atom from the bound rotation/vibration of the molecule
(a two-dimensional Morse oscillator).
This observation will be of crucial importance concerning the
topological properties of the system.

\section{Gate objects in higher dimensions}

For an object to act as a gate in a scattering process, it must
have manifolds of codimension one in phase space.
In general, time-independent Hamiltonian systems with 
$n$ degrees of freedom can be represented in Poincar\'e 
sections of dimension $D = 2n-2$.
Because of time reversal symmetry, hyperbolic orbits
have stable and unstable manifolds of equal dimensionality:
$d=n-1$.
Thus a stable manifold of a hyperbolic periodic orbit 
is of codimension one in
the Poincar\'e section for $n=2$ only. 
Consequently, as soon
as $n > 2$, the periodic orbits cannot anymore act as gates.

However, it has been pointed out by Wiggins~\cite{wig1} 
that in a three-dimensional configuration space,
a two-dimensional invariant subset which is
hyperbolic in the perpendicular direction
have codimension one stable and unstable manifolds in the
phase space of the three-degree-of-freedom system. 
Such objects
can act as gates to a scattering process, 
provided their location in phase space makes it possible.
In scattering systems like ours that can separate into a subsystem 
with $n-1=2$ degrees of freedom plus $1$ translational degree of
freedom, such a $d=2$ dimensional invariant object naturally exists: 
it consists of the lower-dimensional subsystem plus the 
outgoing particle standing infinitely far away from it.
In fact, it has been shown by Toda \cite{toda}
that in a planar atom-diatom 
collision the dynamics of the molecule and the third atom at rest
defines an object in the four-dimensional Poincar\'e section with 
three-dimensional stable and unstable manifolds.

We show in the following that these manifolds define an invariant
set for the scattering process that is superior in dimension to the 
set defined by the typical---i.e., three-body---periodic orbits. 
In this sense this larger set
is the natural generalization of the usual chaotic invariant set of
two-degree-of-freedom scattering systems.

\section{Invariant sets and their dimensions}

By definition, the intersection points of the stable and unstable
manifolds of an invariant object define an invariant set for the
dynamics.
In our scattering model, the gate object in configuration space is
a two-dimensional subset (an annulus) at the ``end'' 
of the outgoing channel.
In the Poincar\'e section, the gate appears as a 2-d object too.
In the remaining two dimensions in its vicinity, one can 
draw a curve from each point of the gate so that 
initial conditions on the curve will lead to asymptotic 
convergence to the orbit started from the point on the gate.
The stable manifold of the gate is then
the collection of these curves forming locally a three-dimensional object.

Because of the global stretching and folding generated by the 
nonlinear dynamics of the system, 
this manifold also has a fractal structure in the
direction perpendicular to it, so globally it has a total dimension 
$d_g = 3 + \delta $  with $0 < \delta < 1$ being the partial fractal 
dimension in the perpendicular direction.
Because of time reversal symmetry, the unstable manifold must have
the same dimension.
Their intersections define an invariant set with a dimension
\begin{equation}
      d_{sg} = 2 d_g - D = 2 + 2 \delta , 
\end{equation}
where $D = 4$ is the
dimension of the Poincar\'e section (the embedding dimension), and the 
subscript ``sg'' indicates the invariant {\em set} generated 
by the {\em gate}.
In contrast, if we consider an inner periodic orbit 
(i.e.\, one which is not part of the gate) 
with locally 2-d stable and unstable manifolds 
that have fractal properties in two
other directions characterized by partial fractal dimensions 
$\epsilon_1$ and $\epsilon_2$,
then the invariant set defined by these 
$d_p = 2 + \epsilon_1 + \epsilon_2 $ dimensional
manifolds have a total dimension
\begin{equation}
      d_{sp} = 2 d_p - D = 2 ( \epsilon_1 + \epsilon_2 ) .
\end{equation}

In order to compare the dimensionalities of these two different invariant
sets, 
we first consider the partial dimensions $\delta$, $\epsilon_1$ and
$\epsilon_2$ of the invariant manifolds.
Since the stable manifolds of two different objects cannot cross, 
the 2-d branches of stable manifolds of periodic orbits must run 
locally parallel to one another and, in particular, 
along the 3-d branches of the stable manifold of the gate object.
Moreover, they are on one side of the stable gate manifold. 
This is because if a trajectory is to approach a certain 
periodic orbit from a given initial point in phase space, 
then it must avoid escape until it reaches the
vicinity of the periodic orbit. 
This can happen only if the starting
point is on the ``right'' side of the 3-d stable branch of the gate.
Therefore the stable branch of the periodic orbit going through 
that point must follow the foldings of the gate manifold, 
so one of the two directions
along which the periodic orbit manifolds show fractality must coincide
locally with the fractal direction of the gate manifold.
This implies that the corresponding partial fractal dimension---say, 
$\epsilon_1$---is equal to $\delta$.
The spatial relationship of the gate manifolds and those of 
the inner periodic orbits is illustrated schematically 
in Fig.~\ref{fig-schem}.

As an immediate consequence, we obtain that $ d_{sg} > d_{sp} $, 
i.e., the invariant set generated by
the gate manifolds is of higher dimension than the invariant set 
belonging to the inner periodic orbits.
In other words, we have a situation where the inner periodic orbits 
are nowhere dense in the set of nonescaping orbits.
This is in sharp contrast to 
two-degree-of-freedom open Hamiltonian systems, where the inner
periodic orbits are everywhere dense in the invariant 
phase space sets.
Our example demonstrates that the denseness of periodic orbits 
in invariant sets of higher dimensional
systems cannot be taken for granted without further considerations.
In our case, without the periodic orbits of the gate itself,
considered ``trivial'' from the point of view of the full dynamics
of the system, there is no dense set of periodic orbits for the 
larger invariant set.

\section{Stable and unstable manifolds in the triple Morse-system}

Following the approach of Chen et al.\ (Ref.~\cite{ding}), 
we can represent the
stable and unstable manifolds and the invariant sets on planar (2-d) 
sections of the four-dimensional phase space.
For this purpose, we chose initial conditions on the Poincar\'e section
with two fixed restrictions and two free parameters.
In this paper, the Poincar\'e section is defined by $y=0$ (with $v_y > 0$) 
while the 
particular restrictions in the initial conditions are $ x_0 + z_0 = 0$ 
and $v_{0x} + v_{0z} = 0$.
We plotted each initial point on the $(z_0, v_{0z})$ plane by using a 
colour coding according to the exit channel the trajectory starting 
from that point eventually took; the result is shown 
in Fig.~\ref{fig-stab}. 

It is clear from this picture that all three colours form compact regions 
with smooth boundaries.
The boundary of a single-coloured region 
consists of trajectories with vanishing translational
kinetic energy at the ``end'' of the channel,
i.e., those asymptotic to the corresponding gate object.
In other words, the smooth curves obtained as the boundaries of
regions of a given colour are just the planar sections of the
stable manifold of the gate closing the channel associated 
with that colour.
On the outer side of these curves,
smaller, differently coloured regions can be found, accumulating
on the boundaries in a fractal manner.
Similar pictures have been presented in Ref.~\cite{boyd} for the 
gravitational three-body problem.

In fact, the stable manifold curves of Fig.~\ref{fig-stab} 
are an example
of Wada boundaries \cite{wada1}, i.e.\ fractal boundary sets 
where all three colours are present in any neighbourhood of the boundary.
It is worth noting that in chaotic scattering processes with more than 
two distinct possibilities for the outgoing motion, Wada boundaries 
are typical \cite{wada2}, but such objects can also appear, e.g., as 
physical boundaries between dyes of different colour poured into open 
hydrodynamical flows \cite{hydro-wada}.

In order to obtain the planar sections of the 
unstable manifolds of the gate objects, 
one should follow the trajectories from the same initial
conditions backward in time. 
However, our initial conditions lie in a symmetry plane of the potential
(T-shape configurations), so following a trajectory backward from an 
initial condition $(z_0, v_{0z})$ is equivalent to following the 
forward-time trajectory starting from the point $(z_0, -v_{0z})$.
Thus the unstable manifold curves in our planar section can be
obtained by mirroring the stable manifold curves with respect to
the $z_0$ axis in Fig.~\ref{fig-stab}.

The sets of stable and unstable manifold curves cross each other,
and the crossing points form the planar section of the invariant set
associated with the gates.
In the case of Fig.~\ref{fig-stab}, 
we can plot an approximation of the cross section points by considering
a colour pixel as a cross section point if both itself and its mirror image 
on the other side of the $z_0$ axis have at least one neighbour cell 
of different color.
The result is shown on Fig.~\ref{fig-cross}; 
the fractal nature of the plot is indicated by a blowup.
According to the considerations of the dimensions in the previous
section,
each point in this plot represents a smooth two-dimensional object
in the total invariant set embedded in the
four-dimensional phase space of the Poincar\'e map.
In other words, our plot captures the fractal part of the invariant
set associated with the gate, having a fractal dimension of
$2\delta$.

It is important to stress that since the crossing of two 2-d objects
in a 4-d space is generic,
we can expect that a similar plot of the gate invariant set can be
obtained in all similar systems with any value of $\delta$.
On the other hand, the invariant set associated with the inner
periodic orbits may or may not produce points in a planar section
depending on whether 
$\epsilon_1 + \epsilon_2$ is larger or smaller than 1;
see the discussion of this aspect in Ref.~\cite{ding}.

\section{Linear sections and scaling properties}

One of the convenient characteristics of two-degree-of-freedom chaotic
scattering
is that its scaling properties can be studied through one-dimensional
sets of initial conditions.
However, as has been shown in Ref.~\cite{ding}, 
this property may or may not be true in general hyperbolic 
chaotic scattering with 3 degrees of freedom, 
depending on the value of the fractal
dimension of the invariant set and its stable manifold.
In this paper, we have shown evidence that in 
three-degree-of-freedom scattering
systems with the asymptotic separation of one degree of freedom,
the feasibility of the one-dimensional description is restored.
The reason for this is that in such systems the crossing of the 
3-d stable manifolds with a line in the 4-d phase space is a generic
property.
This means that typically any 1-d family of initial conditions would
lead to scattering functions with a fractal set of singularities, as
in two-degree-of-freedom chaotic scattering.
These singularities are the fingerprints of the fractal structure
of the stable manifold of the gate.

To check this point, we have produced plots of 
the scattering time for various linear sets of initial conditions 
in our triple Morse model.
They all showed the typical singular behaviour well known 
from two-degree-of-freedom chaotic scattering examples.
We also analyzed the singularities to determine 
the {\em uncertainty exponent} $\alpha_u$, which is related 
to the (partial) fractal dimension as $\delta = 1 - \alpha_u$ \cite{dimu}.
The uncertainty exponent can be measured by choosing pairs of
initial conditions with a separation of $\varepsilon$ along the line:
the rate $p(\varepsilon)$ of pairs where the two orbits escape along 
different channels scales as $p(\varepsilon) \sim \varepsilon^{\alpha_u}$.
We have obtained a value of $0.12$ for $\alpha_u$,
which gives a fractal dimension $\delta = 0.88$ for the stable
manifold of the gates.

It is worth noting that although we measured the fractal dimension
of the stable manifold of the gates, this scaling behaviour 
cannot originate from the gates themselves since they 
are only marginally unstable due to the behaviour of the Morse potential:
The marginal instability of the gates should lead to an
asymptotic fractal dimension value of 1.
However, if our statistics is based on 
moderately long scattering orbits,
we can still observe a scaling region associated with an 
apparent fractal dimension which is 
lower than 1 as if there were only hyperbolic orbits in the system.
In our model, the only truly hyperbolic orbits are the inner periodic
orbits, so the observed scaling can only be produced by them.
This indicates that although these orbits form an invariant set
inferior in dimension to the invariant set of the gate,
they can still dictate the fractal scaling properties, based on 
moderately long orbits, of the larger invariant set.

\section{Conclusions}

We have shown in the example of a simple model that for a 
large class of three-degree-of-freedom chaotic scattering systems, 
the asymptotic
separation of a translational degree of freedom leads to a gate
object regulating the escape process, and that
the gate itself then defines an invariant set
of larger dimension than the one generated by the hyperbolic 
inner periodic orbits.
In fact, although the invariant set associated with these 
periodic orbits is part of the invariant set associated with the 
gate,
the inner periodic orbits are nowhere dense in the larger invariant set
of the gate.
The usual assumption that periodic orbits are dense in an invariant set
is valid in these cases only if we consider also the periodic orbits
in the gate, although they, as ``trivial'' ones, are not readily 
associated with the ``true'' three-body dynamics of the system.
Actually, most of the points of the larger invariant set
are either periodic orbits of the gate or asymptotic to them, leaving
zero measure to scattering trajectories asymptotic to inner periodic 
orbits.

We have also demonstrated, however, that the inner periodic orbits
can still determine the scaling properties of the larger invariant
set due to the fact that their manifolds run locally alongside the
manifolds of the gate object.
This leaves open the question concerning which sets of periodic orbits
are to be taken into account in descriptions based on periodic orbit
theory in larger dimensional systems: 
in general, we cannot exclude the existence of cases when
the gate periodic orbits have dominating contributions to sums
involving all periodic orbits of the system.

Although we treated only one example in three-degree-of-freedom 
chaotic scattering,
our findings can be generalized to any Hamiltonian problem described
by a four-dimensional Poincar\'e map with a suitable 
two-dimensional invariant subspace. 
Another possible way of extension can be considering systems with
more than 3 degrees of freedom: if there is an asymptotic separation of 
only one degree of freedom from the rest, then in principle gate
objects can be defined in an analogous way, and the topological
consequences can be similar too. 
An obvious example is the full spatial dynamics of a three-body
collision that can be reduced to four nontrivial degrees of freedom.

\section*{Acknowledgments}

We thank A. Bringer, G. Haller, E. Ott, L. Rondoni, T. T\'el, 
and J. A. Yorke for insightful discussions. 
One of the authors (Z. K.) is grateful for the kind hospitality of
Laboratoire Spectrom\'etrie Physique, 
which is ``Unit\'e Mixte CNRS/Universit\'e, No.\ 5588'', 
and the financial support by R\'egion Rh\^one-Alpes, program TEMPRA.
This work has been partially supported by the 
Hungarian Scientific Research Foundation 
             (Grant Nos.\ OTKA F17166, 17493 and 19483),
the US--Hungarian Science and Technology Joint Fund
             (Project Nos.\ JF 286 and 501), and
the German--Hungarian Cooperation in Science 
             (Project No.\ D125/95).


\begin{figure}
\caption{$K=0$ surfaces in the abstract representation for $E=2.4$.
         Because of the mirror symmetry of the potential, only the 
         $z \ge 0$ halves are shown.
         The trajectories must stay in the domain between the red and green 
         surfaces.
         The $z=0$ contours on the base are the $K=0$ curves for collinear
         configurations.
         The labels next to the exit channels give the colour coding 
         used in Fig.~\ref{fig-stab} for the outcome of the scattering 
         process.}
\label{fig-cont}
\end{figure}

\begin{figure}
\caption{Schematic picture of the stable manifold of the gate object
         and those of the inner periodic orbits. 
         In a two-dimensional slice of the 
         four-dimensional phase space, the 2-d stable manifolds of 
         periodic orbits appear as dots forming a double Cantor set structure. 
         The 3-d stable manifold of the gate object appears 
         in the same section as a continuous line folded by the dynamics
         so that it has fractal structure in one direction.
         The partial fractal dimension $\delta$ of the folded line 
         agrees with the partial fractal dimension $\epsilon_1$ of the 
         double Cantor set along the same direction.
         This picture also indicates that in the vicinity of a periodic 
         orbit manifold, one can always find 
         points from the stable manifold of the gate, 
         while the opposite is not true.
         The binary organization of the Cantor set and the folding structure
         is chosen only for simplicity.}
\label{fig-schem}
\end{figure}

\begin{figure}
\caption{Initial conditions from the 2-d subspace of the Poincar\'e section
         coloured according to the exit channel of the corresponding 
         trajectory.
         The boundary curves of the single-colour regions are the slices 
         of the stable manifold of the gate object.}
\label{fig-stab}
\end{figure}

\begin{figure}
\caption{Approximation of the planar cross section of the invariant set 
         defined by the gate manifolds (a).
         The picture shows the signs of a fractal structure indicated by
         the blowup (b) of a small region that looks homogeneous on a lower
         resolution.}
\label{fig-cross}
\end{figure}

\end{document}